\documentclass[preprintnumbers,nofootinbib]{revtex4}
\usepackage{graphicx}
\usepackage{dcolumn}
\usepackage{bm}
\usepackage{epsfig,amsmath}
\usepackage{amssymb}
\usepackage{subfigure}
\usepackage{float}
\usepackage{ulem}
\usepackage[usenames,dvipsnames]{color}
\usepackage[pagebackref=false, colorlinks=true]{hyperref}
\usepackage[labelfont=bf]{caption}
\definecolor{redish}{rgb}{0.7,0.2,0.0}  
\definecolor{bluish}{rgb}{0.2,0.5,0.8}

\makeatletter
\renewcommand{\fnum@figure}{Figure. \thefigure}
\makeatother

\hypersetup{linkcolor=redish,          
                  citecolor=blue,        
                  filecolor=magenta,      
                  urlcolor=bluish}          

\DeclareFontFamily{U}{rsfs}{}         
\DeclareFontShape{U}{rsfs}{m}{n}{<5> rsfs5 <6><7> rsfs7          %
  <8><9><10><10.95><12><14.4><17.28><20.74><24.88> rsfs10}{}     %
\DeclareMathAlphabet{\mathfs}{U}{rsfs}{m}{n}

\def \O{\Omega}

\def \f{\frac}

\def \o{\omega}

\def \a{\alpha}
\def \b{\beta}

\def \S{\Sigma}

\def \p{\partial}

\def \a{\alpha}
\def \L{\Lambda}
\def \l{\lambda}
\def \th{\theta}

\def \ra{\rightarrow}
\def \D{\Delta} 

\begin{document}

\title{Magnetic Penrose process in the magnetized Kerr spacetime}

\author{Chandrachur Chakraborty}
\email{chandrachur.c@manipal.edu} 
\affiliation{Manipal Centre for Natural Sciences, Manipal Academy of Higher Education, Manipal 576104, India}
\author{Parth Patil}
\affiliation{Manipal Centre for Natural Sciences, Manipal Academy of Higher Education, Manipal 576104, India}
\author{G. Akash}
\affiliation{Department of Astronomy, Astrophysics and Space Engineering, Indian Institute of Technology,
Indore 453552, India}

\begin{abstract}
It is well-established that the magnetic Penrose process (MPP) could be highly efficient (efficiency can even exceed $100\%$) for extracting the energy from a Kerr black hole, if it is immersed in a mG order magnetic field. Considering the exact solution of the magnetized Kerr spacetime, here we derive the exact expression of efficiency ($\eta_{\rm MPP}$) for MPP, which is valid for both the Kerr black hole (BH) as well as Kerr superspinar (SS), and also from the weak magnetic field to an ultra-strong magnetic field $(B)$ which can even distort the original Kerr geometry. We show that although the value of $\eta_{\rm MPP}$ increases upto a certain value of ultra-strong magnetic field ($B_p$), it decreases to zero for $B > B_p$, in case of the Kerr BHs. On the other hand, $\eta_{\rm MPP}$ shows the opposite behavior in case of the Kerr SSs. One intriguing feature that emerges is, $\eta_{\rm MPP}$ acquires the maximum value for the Kerr parameter $a_* \approx 0.786$ (unlike $a_*=1$ for the ordinary PP), decreases for the range $0.786 < a_* \leq 1$, and reaches to $20.7\%$ for $a_*=1$ with a few limitations. This indicates that the BH starts to expel the effect of magnetic field for $a_* > 0.786$, and is fully expelled from the extremal Kerr BH due to the gravitational Meissner effect. 
As a special case of MPP, we also study the ordinary Penrose process (PP) for the magnetized Kerr spacetime. We show that the efficiency of PP decreases with increasing the magnetic field for the Kerr BH. In case of the Kerr SS, the efficiency of PP decreases from $10^3\%$ to $0$ for increasing the value of magnetic field from $0$ to a specific value of ultra-strong magnetic field. Thus, the MPP for Kerr BHs, Kerr SSs and the ordinary PP for Kerr SSs can be superefficient for the astrophysical applications to powering engines in the high-energy sources like active galactic nuclei and quasars, in the weak magnetic fields. Our strong magnetic field result of MPP could be important to the primordial BHs in the early Universe immersed in the primordial magnetic fields, and to the transmuted BHs which are formed by collapsing and/or by merging of the magnetized progenitors. It is almost impossible to extract the energy from a BH (SS) through MPP (PP) in the ultra-strong magnetic fields.
\end{abstract}

\maketitle

\section{\label{intro}Introduction}

Penrose process (PP) is a mechanism by which the rotational energy of a Kerr black hole (BH) can be extracted \cite{Penrose69}. As proposed by Roger Penrose in 1969, the energy extraction is achieved when a parent particle gets split into two fragments in the vicinity of the BH. In the idealistic scenario the split would occur very close to the event horizon. Out of the two fragments one would attain the negative energy orbits inside the ergoregion and one would escape to infinity. The escaping particle in this scenario would carry more energy than its parent particle, because of attainment of negative energy state by other particle. For a successful energy extraction it is necessary that the split should occur in the ergoregion, as there only negative energy states are available. Maximum $29\%$ \cite{dh} of the BH energy can be extracted in this fashion from a maximally rotating BH. Further investigations were conducted to test its astrophysical viability \cite{bpt,Wald74a,Wald74b}, the conclusions of which were not in favor of PP. The poor efficiency of only $20.7\%$ \cite{dh} for a maximally rotating BH, and the condition that the relative velocity between two fragments should be of order $c/2$, \cite{bpt,Wald74a,Wald74b} become the Achilles heel of PP. Note that the efficiency of PP depends only on the geometry of the spacetime, and, it does not depend on the mass of the BH.

In 1985, the magnetic version of the PP was proposed by Wagh et al. \cite{Wagh85}, which is known as the magnetic Penrose process (MPP). This is almost similar to the PP with two major modifications. First one is, the Kerr BH should be surrounded by the magnetic fields. Second one is, the neutral particle should be split into two charged fragments. One charged particle attains the negative energy, and falls into the BH. Another charged particle escapes to infinity with energy exceeding the energy of the incident particle to conserve the total energy \cite{Wagh85}. Later, it was shown \cite{bhat, ps86} that the efficiency of MPP could even exceed $100\%$ if a stellar mass Kerr BH is surrounded by a mG order magnetic fields \cite{dh}. Note that the efficiency of MPP does not only depend on the geometry of the spacetime, but it also depends on the non-vanishing components of the four-potential $A_{\mu}$ of the electromagnetic fields and some other parameters. Remarkably, the condition of the relative velocity between the two fragments $c/2$ (where $c$ is the speed of light in vacuum) for PP \cite{bpt,Wald74a,Wald74b}, is nicely circumvented in MPP. This is because the required energy could now come from the electromagnetic fields leaving the relative velocity between the two fragments to be free \cite{dh}. However, the formalism of MPP was developed \cite{Wagh85,ps86,bhat} based on the ordinary Kerr BH assuming the weak magnetic field, and the components of $A_{\mu}$ was presumed to be the linear order in $B$ \cite{dh}. In another work, it is recently proposed \cite{dyson} that the accretion can happen in a superradiant manner, and the angular momentum can be extracted from the dyonic Kerr-Newman BH via Penrose process in an enhanced rate due to the dipolar electric field created by the rotating magnetic charge of the BH.

In reality, the magnetic field plays an important role to explain the various
astrophysical phenomena, e.g., the magnetohydrodynamics (MHD) simulation for the accretion mechanism \cite{gam}, polarization of the BH shadow \cite{eht7, eht8, c22}, gravitational Larmor precession \cite{glp, gp}, gravitational Meissner effect \cite{bicak}, MPP \cite{dh} etc. Although the magnetic Penrose process was proposed assuming the magnetic field to be asymptotically uniform \cite{dh}, the Blandford-Znajek (BZ) mechanism \cite{bz} is generally considered for the MHD simulation. Due to the lacking of the proper measurements of the exact shapes of the magnetic field configurations around a collapsed object, many other numerical techniques are used to show the strong connections between the shape of magnetosphere and the characteristics of accretion mechanism \cite{pun, mei, shld}. To derive the exact solution of MPP, here we consider the Ernst \cite{er} and/or Wald \cite{wald} solution of the magnetized Kerr spacetime as this is the exact electrovacuum solution of the Einstein-Maxwell equation following \cite{c22}. However, there was some drawback of the Ernst solution \cite{c22}, which was removed by \cite{wild} in order to obtain a physically meaningful solution \cite{ag2}. Later it was applied  to observe the magnetic precession \cite{ag3} in BH systems with magnetized accretion disks, the gravitational Faraday rotation \cite{c22} etc. Note that the gravitational energy is generally much greater than the electromagnetic energy, but they are comparable if the strength of the magnetic field $(B)$ around a collapsed object with mass $M$ is the order of \cite{ag2, gp}
\begin{eqnarray}
 B \simeq B_{\rm max} \sim 2.4 \times 10^{19} \f{M_{\odot}}{M} ~{\rm Gauss}
 \label{bmax}
\end{eqnarray}
where $M_{\odot}$ is the solar mass. 
The strength of the magnetic field around a BH is considered to be much smaller than the value of $B_{\rm max}$ (i.e., $B << B_{\rm max}$ as considered in the original formulation of MPP \cite{Wagh85, ps86, dh}) but the investigations suggest that the surrounding spacetimes around a BH could be highly distorted for $B \sim B_{\rm max}$. Thus, the magnetic field is important as a background field testing the geometry around a collapsed object \cite{sha}. That is the reason, we do not make any assumption and use any approximation on the field intensity of the magnetic field.

In this paper, we consider the more general magnetized Kerr spacetime which is the exact electrovaccum solution of the Einstein-Maxwell equation with exact components of the four-potential $A_{\mu}$ of the electromagnetic fields, and derive the exact expression of the efficiency of the MPP, without making any assumption on the intensity of the magnetic fields, value of the Kerr parameter ($a_*$) and $A_{\mu}$. Therefore, our result is valid for the (magnetized) Kerr BH ($0 < a_* \leq 1$) as well as Kerr SS ($a_* > 1$) \cite{gh, ckj, ckp}, and also from the weak magnetic fields to the ultra-strong magnetic fields which can even distort the original Kerr geometry. Our exact result of the efficiency of MPP in the magnetized Kerr spacetime reduces to the result obtained in \cite{dh, Wagh85, ps86} for a Kerr BH immersed in a weak magnetic field. Similarly, our exact result for the efficiency of PP in the magnetized Kerr spacetime reduces to the result
obtained in \cite{Penrose69} for an ordinary (unmagnetized) Kerr BH.
The paper is organized as follows. The formalism of MPP and ordinary PP in a general stationary and axisymmetric spacetime are discussed in Sec. \ref{sec2}. Sec. \ref{sec3} is devoted to describe the magnetized Kerr spacetime and the properties of its surrounding electromagnetic fields. We derive the exact expressions of efficiencies for MPP and PP in the magnetized Kerr spacetime, and discuss our result for the magnetized Kerr BHs and Kerr SSs in Sec. \ref{sec4}. Finally we summarize and discuss the limitation of the formulation in Sec. \ref{sec6}.

\section{\label{sec2}Energy extraction through the magnetic Penrose Process from a stationary and axisymmetric spacetime}

We consider a stationary and axisymmetric spacetime with the line element 
\begin{eqnarray}
 ds^2=g_{tt}dt^2+2g_{t\phi} dt d\phi+g_{\phi\phi} d\phi^2+g_{rr}dr^2+g_{\th\th}d\th^2.
 \label{gen}
\end{eqnarray}
Eq. (\ref{gen}) should violate the time reflection symmetry ($t \ra -t$) to satisfy the stationarity condition (see \cite{wald} for details). In addition, the spacetime has to be symmetric about an axis. $g_{tt}$ could vanish for some specific values of $r$, and the biggest (smallest) root among them is $r \equiv r_e$ ($r_i$). Similarly, $g_{rr}$ diverges for some specific values of $r$, and the biggest root among them is $r \equiv R_h$. The bounded region between $r_e$ and $R_h$ is known as the (outer) ergoregion, as the extraction of energy might be possible from this particular region by the Penrose process. If $R_h$ is obtained as imaginary (i.e., the event horizon does not exist) for any special case, the bounded region between $r_e$ and $r_i$ is known as the ergoregion. For example, the ergoregion is defined by the bounded region between the inner and outer ergoradii in case of the Kerr naked singularity (see \cite{ckj, ckp} for details). If $r_e$ is equal to $R_h$ for all values of $\th$, the ergoregion does not arise for that specific case, and the energy extraction is impossible through the Penrose process from that spacetime (e.g., Schwarzschild BH, Taub-NUT BH \cite{cm}).

Let us consider an axisymmetric and stationary electromagnetic field superposed on the above-mentioned geometry (Eq. \ref{gen}) that could be described by the 4-potential $A_{\mu} \equiv (A_t, 0, 0, A_{\phi})$. The Lagrangian ($\mathcal{L}$) of a test particle of mass $m$ and charge $q$ (with the charge to mass ratio $\l=q/m$) moving in the above-mentioned electromagnetic field is expressed as
\begin{equation}
\mathcal{L} = \frac{1}{2}m g_{\mu\nu}\Dot{x}^{\mu}\Dot{x}^{\nu} + qA_{\mu}\Dot{x}^{\mu}
\end{equation}\\
where $\Dot{x}^{\mu} (\equiv U^{\mu})$ is the 4-velocity of the particle and the dot represents the differentiation with respect to the proper time $(\tau)$ of the
corresponding particle. Since the metric and the electromagnetic field both are stationary and axisymmetric, the $t$ and $\phi$ components of the generalized momentum ($P_{\mu}$) of the particle are conserved. Thus, we can write the following integrals of motion \cite{ps86}
\begin{eqnarray}
   P_t &=& \frac{\partial\mathcal{L}}{\partial\Dot{t}} = m U_{t}+qA_{t} = -m\mathcal{E} 
\\
     P_{\phi} &=& \frac{\partial\mathcal{L}}{\partial\Dot{\phi}} = mU_{\phi}+qA_{\phi} = ml
\end{eqnarray}
or, equivalently we can define
\begin{eqnarray}
   U_{t}=-(\mathcal{E}+\l A_{t}) &=& -E 
   \\
 U_{\phi}=l-\l A_{\phi} &=& L
\end{eqnarray}
where $\mathcal{E}$ and $l$ are the energy and angular momentum per unit mass of the test particle, respectively.

In the Penrose process, a neutral particle of mass $m_{1}$ incident into a collapsed object is supposed to split into two fragments, e.g., particle $2$ of mass $m_2$ and particle $3$ of mass $m_3$. At the point of split the energy ($m\mathcal{E}$), the angular momentum ($ml$), the linear momentum ($m\dot{r}$) and the charge ($m\l$) are conserved \cite{ps86, dh}, respectively, i.e.,
\begin{eqnarray}\nonumber
     \mathcal{E}_{1} &=& m_2 \mathcal{E}_{2}+m_3 \mathcal{E}_{3},
     \label{e}
\\ \nonumber
    l_{1} &=& m_2 l_{2}+ m_3 l_{3},
    \\ \nonumber
    \dot{r}_{1} &=& m_2 \dot{r}_{2}+ m_3 \dot{r}_{3},
    \\ 
      \l_{1} &=& m_2 \l_{2}+m_3 \l_{3}. 
      \label{elc}
\end{eqnarray}
In Eq. (\ref{elc}), we set $m_1=1$ without the loss of generality, that means we are measuring $m_2$ and $m_3$ in terms of $m_1$ \cite{ps86}. The subscripts $1, 2, 3$ represent the corresponding physical quantities of the incident particle and two fragments respectively. In addition we restrict the sum
of masses of the fragments after split to $m_2+m_3 \leq 1$.
Let particle 2 attains the negative energy \cite{bhat} $\mathcal{E}_{2}<0$, and falls into the event horizon, while particle $3$ escapes to infinity with higher energy $\mathcal{E}_{3} ~(~ \mathcal{E}_{3} > \mathcal{E}_1)$ to conserve the total energy. To achieve the maximum efficiency for the Penrose process, one can set $\dot{r}_{2}=0$ which implies $\dot{r}_{1} =m_3 \dot{r}_{3}$. Thereby, no kinetic energy
is lost through the particle $2$. In this way, the energy of a collapsed object could be extracted. However, the motion of a charged particle of unit mass is bounded by the following effective potential \cite{ps86}
\begin{eqnarray}
 V=-\l A_t+ \o L + \left[(-\Psi) \left(\f{L^2}{g_{\phi\phi}}+1 \right) \right]^{1/2}
\end{eqnarray}
where $\Psi=g_{tt}+\o g_{t\phi} < 0$, and $\o=-g_{t\phi}/g_{\phi\phi}=d\phi/dt$ is the angular velocity of a locally nonrotating observer.

At the point of splitting of the neutral particle, the 4-momenta $P_i~ (i=1,2,3)$ of all the three particles are timelike, and one can define the 4-velocity vector as ${\bf U}=\dot{t}(1,v,0,\O)$, where $\dot{t}=dt/d\tau$, $v=dr/dt$ and $\O=d\phi/dt$. Therefore, one can write
\begin{eqnarray}
 {\bf U}. {\bf \xi} =-(\mathcal{E}+\l A_t)=-E
\end{eqnarray}
where ${\bf \xi} \equiv (\p_t)$ is the timelike Killing vector field, which yields
\begin{eqnarray}
 \dot{t}=-E/X
 \label{ex}
\end{eqnarray}
with $X=g_{tt}+\O g_{t\phi}$. Using the 4-velocity  ${\bf U}$
with ${\bf U}.{\bf U}=-1$, one can write
\begin{eqnarray}
 g_{tt}+2g_{t\phi}\O+g_{\phi\phi}\O^2+g_{rr} v^2=-(X/E)^2 \leq 0.
\end{eqnarray}
There is a limit on $\O$ (say, $\O_{\pm}^{\rm gen}$),
\begin{eqnarray}
 \O_{\pm}^{\rm gen}= \o \pm \f{1}{g_{\phi\phi}}\sqrt{-\psi -g_{\phi\phi} g_{rr} v^2}
 \label{opm}
\end{eqnarray}
(where $\psi= g_{tt}g_{\phi\phi} - g_{t\phi}^2$) tending to which, ${\bf U}$ tends to a null vector. It indicates that the allowed values of $\O$ at any fixed $(r, \th)$ are
$ \O_+^{\rm gen} < \O < \O_-^{\rm gen}$ \cite{ckp}.
This follows from Eq. (\ref{elc})
\begin{eqnarray}\nonumber
  \dot{\phi}_{1} &=& m_2 \dot{\phi}_{2}+ m_3 \dot{\phi}_{3},
    \\
    \dot{r}_{1} &=& m_2 \dot{r}_{2}+ m_3 \dot{r}_{3},
\end{eqnarray}
or, equivalently
\begin{eqnarray}\nonumber
  \O_{1} \dot{t}_{1} &=& m_2 \O_{2} \dot{t}_{2}+ m_3 \O_{3} \dot{t}_3,
    \\
    v_1 \dot{t}_{1} &=& m_2 v_2 \dot{t}_{2}+ m_3 v_3 \dot{t}_{3}.
    \label{vdot}
\end{eqnarray}
Let us now assume that a neutral particle splits, i.e., $\l_{1} \ra 0$. As the astronomical bodies does not possess any charge, this assumption is well justified \cite{ps86}. This requires the incident particle to be uncharged. Using Eq. (\ref{e}), Eq. (\ref{ex}) and Eq. (\ref{vdot}), one can obtain from Eq. (\ref{e}) for $m_3 \mathcal{E}_{3}$:
\begin{eqnarray}
  m_3 \mathcal{E}_{3} &=& \chi \mathcal{E}_{1}-  m_3 \l_3 \mathcal{A}_t
  \label{m3c3}
\end{eqnarray}
where 
\begin{eqnarray}
	\chi &=& \left(\frac{\Omega_{1}-\Omega_{2}}{\Omega_{3}-\Omega_{2}}\right)\frac{X_{3}}{X_{1}}
	\\
	&=& \left( \f{v_1 X_2-v_2 X_1}{v_3 X_2-v_2 X_3} \right)\f{X_{3}}{X_{1}}
	\label{chi}
\end{eqnarray}
with $ X_{i} = g_{tt}+\Omega_{i}g_{t\phi}$. One can also write down 
\begin{eqnarray}
 v_1(\Omega_{3}-\Omega_{2})+v_2 (\Omega_{1}-\Omega_{3}) +v_3 (\Omega_{2}-\Omega_{1})=0
\end{eqnarray}
from Eq. (\ref{chi}), which is considered as the coplanarity condition for the 3-momenta: ${\bf P}_1={\bf P}_2+{\bf P}_3$. Finally, the general expression for efficiency ($\eta^{\rm gen}$) can be written as
\begin{eqnarray}
 \eta^{\rm gen}=\f{ m_3 \mathcal{E}_{3}-\mathcal{E}_{1}}{\mathcal{E}_{1}} = \chi -1 - \f{m_3 \l_3 \mathcal{A}_t}{\mathcal{E}_{1}}
  \label{etag}
\end{eqnarray}
using Eq. (\ref{m3c3}).
It is shown \cite{ps86} that $\chi$ is maximized when all the radial velocities are zero, i.e., $v \ra 0$ in Eq. (\ref{opm}), and $\Omega_{2}=\Omega_{-}$ and $\Omega_{3}=\Omega_{+}$. 
In such a case, $\O_{\pm}^{\rm gen}$ reduce to \cite{bhat, ckp}
\begin{eqnarray}
 \O_{\pm}= \f{1}{g_{\phi\phi}}\left( -g_{t\phi} \pm \sqrt{-\psi} \right)
 \label{opm1}
\end{eqnarray}
and, $\O_1$ reduces to \cite{bhat}
\begin{eqnarray}
 \O_1=\f{-g_{t\phi}(1+g_{tt})+(-\psi(1+g_{tt}))^{1/2}}{g_{t\phi}^2+g_{\phi\phi}}.
\end{eqnarray}
Substituting all those above in Eq. (\ref{etag}), finally we obtain the expression of the efficiency ($\eta$) as \cite{dh}
\begin{eqnarray}
    \eta = \left[ \left(\frac{\Omega_{1}-\Omega_{-}}{\Omega_{+}-\Omega_{-}}\right)\left(\frac{g_{tt}+\Omega_{+}g_{t\phi}}{g_{tt}+\Omega_{1}g_{t\phi}}\right) -1 \right]-\frac{q \mathcal{A}_{t}}{m}
    \label{etaf}
\end{eqnarray}
where $m_3 \l_3=q_3 \equiv q$ that is the charge of the outgoing particle of mass $m$. In Eq. (\ref{etaf}), $\mathcal{E}_{1}$ is taken as $\mathcal{E}_{1} \approx 1$ following \cite{ps86} as the incident particle initially moves nonrelativistically in almost all realistic situations \cite{ps86}. In addition, for the realistic calculation, the last term of Eq. (\ref{etag}) is replaced by $q\mathcal{A}_{t}/m$ in Eq. (\ref{etaf}) following \cite{dh}  (see also \cite{ps86} and the discussion below Eq. \ref{elc}), so that it becomes dimensionless in the geometrized unit ($G=c=1$).
Note that the term in the square bracket of Eq. (\ref{etaf}) is purely a geometric factor which depends only on the metric components, whereas the other term that remains outside of it, is not a geometric factor. It depends on the mass and charge of the outgoing particle, and  $\mathcal{A}_{t}$ including the metric components, which would be cleared as we proceed.
If $q$ and/or $\mathcal{A}_{t}$ vanish, Eq. (\ref{etaf}) gives the efficiency ($\eta \equiv \eta_{\rm PP}$) for the ordinary Penrose process \cite{Penrose69}. On the other hand, if a non-zero $\mathcal{A}_{t}$ arises (with a non-zero $q$) due to the presence of a magnetic field, Eq. (\ref{etaf}) can provide the efficiency ($\eta \equiv \eta_{\rm MPP}$) of the magnetic Penrose process \cite{dh}.

However, now one can easily apply Eq. (\ref{etaf}) to find the efficiency for  various axisymmetric and stationary spacetimes. 
In this paper, we are going to apply it for the Kerr spacetime which is immersed in a magnetic field. Thus, let us first briefly discuss the magnetized Kerr spcatime in the next section.

\section{\label{sec3}Kerr spacetime immersed in a uniform magnetic field}

The exact electrovacuum solution of the Einstein-Maxwell equation for the  magnetized Kerr spacetime is written as \cite{ast, ag1, ag2}
\begin{eqnarray}
 ds^2=\left( -\f{\D}{A}dt^2+\f{dr^2}{\D}+d\th^2 \right)\S|\L|^2+\f{A\sin^2 \th}{\S|\L|^2}\left(|\L_0|^2 d\phi-\varpi dt \right)^2
 \label{Kerrm}
\end{eqnarray}
where 
\begin{eqnarray}
  \D=r^2+a^2-2Mr \, , \,\,\,\,\, \S=r^2+a^2\cos^2\th ,
\end{eqnarray}
\begin{eqnarray}
A=(r^2+a^2)^2-\D a^2 \sin^2\th \, , \,\,\,\,\, \varpi=\f{\a -\b \D}{r^2+a^2}+\frac{3}{4}aM^2B^4.
 \label{ob}
\end{eqnarray}
$M$ and $a$ are the mass and spin parameter of the Kerr spacetime (respectively) which is immersed in an uniform magnetic field $(B)$.
$\L (r,\th)$ is a complex quantity and it has two parts, the real part of $\L$: ${\rm Re}~\L$ and the imaginary part of $\L$: ${\rm Im}~\L$. Thus, one can express it as:
\begin{eqnarray}
 \L \equiv \L (r,\th) &=& {\rm Re} \L + i~{\rm Im} \L \nonumber
 \\
 &=&1+\f{B^2\sin^2\th}{4}\left[\left(r^2+a^2\right)+\f{2a^2Mr\sin^2\th}{\S} \right]-i.\f{aB^2M\cos\th}{2}\left(3-\cos^2\th+\f{a^2\sin^4\th}{\S}\right) \nonumber
 \\
 \label{lambda}
\end{eqnarray}
where $i~(\equiv \sqrt{-1})$ represents the imaginary unit. In the expression of $\varpi$ (Eq. \ref{ob}), 
\begin{eqnarray}
 \a &=& a(1-a^2M^2B^4) 
 \\
 {\rm and,} && \nonumber
 \\
 \b &=& \f{a\S}{A}+\f{aMB^4}{16}
 \left(-8r\cos^2\th(3-\cos^2 \th)-6r\sin^4\th+\f{2a^2\sin^6\th}{A}[2Ma^2+r(a^2+r^2)] \right. \nonumber
 \\
&+& \left. 
\f{4Ma^2\cos^2\th}{A}\left[(r^2+a^2) (3-\cos^2 \th)^2-4a^2\sin^2\th \right] \right).
\end{eqnarray}
The term
\begin{equation}
|\Lambda_{0}|^{2} \equiv |\Lambda(r,0)|^{2}=1+a^{2}M^{2}B^{4}
\end{equation}
is introduced in the metric (Eq. \ref{Kerrm}) to remove the conical singularities on the polar axis \cite{ag1, ag2, c22}.
Eq. (\ref{Kerrm}) shows that the event horizon of the magnetized Kerr black hole remains the same as that of a Kerr black hole, i.e.,
\begin{equation}
    r_{\pm}=M(1 \pm \sqrt{1-a_*^{2}})
    \label{hor}
\end{equation}
where $a_*=a/M$. This indicates that the magnetic field $B$ does not have any effect on the event horizon.
Eq. (\ref{hor}) also reveals that the radius ($R_h \equiv r_+$) of the event horizon $R_h =M(1+\sqrt{1-a_*^2})$ becomes imaginary for $a_* > 1$. This indicates that the horizon does not exist, and the ring singularity ($r \ra 0, \th \ra \pi/2$) could be visible from the infinity, in principle \cite{ckp}. However, the quantum gravity effects should be important in the region where the spacetime curvature approaches the Planck scale, i.e., very close to the ring singularity \cite{bambif}. It is also widely accepted that quantum gravity will resolve the singularity resulting in an overspinning object with a boundary at a positive value of $r$ (say, $ r= R_s$), which is referred to as a `superspinar' \cite{gh, ckp}.  The value of $R_s$ could be $r=R_s=10^{-3}M$ following \cite{bambi}. Remarkably, it has recently been proved that the superspinars are stable \cite{nakao} (see also \cite{bambi}).
However, the angular momentum ($J$) of the magnetized Kerr spacetime is affected by $B$ as \cite{ast}
\begin{equation}
    J=aM(1-a^{2}M^{2}B^{4}) \equiv a_*M^2 (1-a_*^2M^4B^4)
    \label{J}
\end{equation}
which vanishes for $a_* \ra 0$ as well as $a_* \ra 1/(B^2M^2)$. This indicates that the angular momentum of an extremal Kerr BH ($a_*=1$) vanishes if it is immersed in the ultra-strong magnetic field $B =M^{-1}$.
For $BM << 1$, one can neglect the second term and write $J \approx a_*M^2$ using Eq. (\ref{J}). The ergoregion is also affected by $B$ for the magnetized Kerr spacetime. One can obtain two positive real roots ($r_{eo}$ and $r_{ei}$ with $r_{eo}>r_{ei}$) as the radius of ergoregion ($r_e$) by solving the following equation,
\begin{equation}
   g_{tt}=-\frac{\Delta \Sigma}{A}|\Lambda|^{2}+\frac{\varpi^{2}A\sin^{2}\theta}{\Sigma |\Lambda|^{2}} =0.
   \label{ergo}
\end{equation}
Although Eq. (\ref{ergo}) cannot be solved analytically, it could be solved numerically to get an impression of the structure of the ergoregion in the presence of magnetic field (see \cite{Gibbons13} for details). 
The solution of Eq. (\ref{ergo}) for $B \ra 0$ reduces to
\begin{eqnarray}
 r_e|_{B \ra 0}=M(1 \pm \sqrt{1-a_*^{2}\cos^2\th}).
\end{eqnarray}
One should note here that although the event horizon does not exist for a superspinar (SS), the ergoregion still exists for it. In case of the Kerr naked singularity, the ergoregion is defined by the bounded region between $r_{ei}$ and $r_{eo}$ (see \cite{ckj} for details). However, in case of the Kerr SS, we do not bother about $r_{ei}$ as it is inside $R_s$, and thus, the ergoregion should be defined by the bounded region between $R_s$ and $r_{eo}$ \cite{gh}. However, its structure would be significantly different with and without magnetic fields. For example, the ergoregion for a weakly magnetized ($BM << 1$) Kerr SS look like a torus, with openings along the axis of rotation \cite{gh} (similar to FIGs. 1 and 2 of \cite{ckp}). Note that the energy from the Kerr BH (Kerr SS) could be extracted from the ergoregion bounded by $R_h$ ($r_{ei}$) and $r_{eo}$ \cite{ckj, ckp}. 

Here the Kerr spacetime is immersed in a magnetic field which has symmetries of stationarity and axial symmetry. As the magnetic field is asymptotically uniform, the non-vanishing components of the four-potential $\mathcal{A}^{\mu}$ can be written as \cite{ast}
\begin{eqnarray}\nonumber
  \mathcal{A}_{\phi} &=& \mathcal{A}_{\phi 0} +    \frac{B |\L_0|^2}{8 \S |\L|^2}  \left\{ a^6 B^2 x^2 \D_x^2 +r^4 \D_x(4+B^2 r^2\D_x) \right.  
   \\ \nonumber
   && \left. +   a^2 r \Big[ 4 B^2 M^2 r x^2 (3-x^2)^2+4M\D_x^2 (2+B^2r^2 \D_x) + r[4-4x^4+B^2r^2(2-3x^2+x^6)] \Big]     \right.    
   \\ 
&& \left.   + a^4 \Big[ 4 x^2 \D_x + B^2 [4 M r \D_x^3 + 4M^2(1+x^2)^2 + r^2(1-3x^4+2x^6) ]\Big] \right\}
\label{aphi}
\end{eqnarray}
and,
\begin{eqnarray}\nonumber
\mathcal{A}_t &=& \mathcal{A}_{t0} - \varpi \left( \frac{\mathcal{A}_{\phi} - \mathcal{A}_{\phi 0} }{|\L_0|^2}\right) 
\\ 
&&  - a M B^3 \left[2r + \frac{1}{A} \left( 4 a^2 M^2 r - r(r^2+a^2) \D \D_x - \frac{1}{4} (r^3 -3 a^2 r + 2 M a^2) \D \D_x^2  \right) \right]
\label{at}
\end{eqnarray}
for $\mathcal{A}=\mathcal{A}_t dt + \mathcal{A}_{\phi}d\phi$,
where $\D_x=(1-x^2)$ and $x=\cos\th$. The two gauge additive constant  $\mathcal{A}_{\phi 0}$ and $\mathcal{A}_{t 0}$ \cite{ast} could be fixed depending on requirement. 
In the linear order in $B$, Eq. (\ref{aphi}) and Eq. (\ref{at}) reduce to
\begin{eqnarray}
 \mathcal{A}_{\phi} = \mathcal{A}_{\phi 0}+\f{AB\sin^2\th}{2\S} + \mathcal{O}(B^2)
 \label{apb}
\end{eqnarray}
and
\begin{eqnarray}
 \mathcal{A}_t = \mathcal{A}_{t0} -\f{aMrB\sin^2\th}{\S}+ \mathcal{O}(B^2)
 \label{atb}
\end{eqnarray}
respectively. For choosing vanishing $\mathcal{A}_{\phi 0}$ and $\mathcal{A}_{t0}$, Eq. (\ref{apb}) and Eq. (\ref{atb}) are not only resemble to Eq. (3.4) of \cite{ag2} but it is also accordance with the gauge $A_t(r \ra \infty)=0$, as described in \cite{ag2, ag1}. This gauge is more suitable to describe the charged particle ergospheres around a magnetized black hole \cite{ag1}.
However, in order to ensure the regularity of the electromagnetic field on the rotation axis of the magnetized Kerr spacetime (Eq. \ref{Kerrm}), $\mathcal{A}_{\phi 0}$ of the exact expression of $\mathcal{A}_{\phi}$ (Eq. \ref{aphi}) has to be gauge fixed such that $\mathcal{A}_{\phi}=0$ at $x=\pm 1$ \cite{ast}. Thus, one  obtains $\mathcal{A}_{\phi 0} = -2a^2 M^2 B^3$ following \cite{ast}.

The magnetic flux ($F_B$) through the upper hemisphere of the event horizon is an important parameter in the theory of the electromagnetic extraction of energy from a Kerr black hole \cite{bz}. The general expression of $F_B$ for the magnetized (Kerr-)Newman spacetime was derived in \cite{ag2}, which reduces to \cite{ag2}
\begin{eqnarray}
 F_B=4\pi B M^2\f{1-a^2B^2}{1+B^2M^2}
\end{eqnarray}
for the magnetized Kerr spacetime. $F_B$ vanishes only for
$a = 1/B$, not necessarily for the extremal Kerr black hole. For the non-charged magnetized solution in the linearized approximation (in $B$) corresponds to the electric charge $Q=-2aMB$, the magnetic flux vanishes (see Eq. 4.5 of \cite{ag2}) for the extremal Kerr black hole, which resembles to the result obtained in \cite{king} (see also \cite{bicak,karas00,bi85} and references therein) within the test field approximation. This result leads to obtain the expression of $\mathcal{A}_t$ in Eq. (2) of \cite{dh}, which vanishes on the horizon of an extremal Kerr black hole. Thereby, to satisfy Eq. (2) of \cite{dh}, we choose $\mathcal{A}_{t0}=aB$ for $\th \ra \pi/2$. Using the concept of comoving potential, we replace $\mathcal{A}_{t}$ with $-\mathcal{A}_{t}$ following \cite{dok}. Considering all those above, finally we obtain
\begin{eqnarray}
\mathcal{A}_t|_{\th \ra \pi/2} = a B \left[
   \f{M}{2} \left(5 B^2 M + 3 B^2 r - \f{a^2 B^2}{r} 
   + \f{2 [4 + B^2 (a^2 - 3 r^2)]}{
      4 r + B^2 [r^3 + a^2 (2 M + r)]}\right)-1\right]
\label{atn}
\end{eqnarray}
for $\th \ra \pi/2$. Eq. (\ref{atn}) reveals that $\mathcal{A}_t$ vanishes at $\th \ra \pi/2$ in the linear order of $B$ for the extremal Kerr BH, as mentioned in \cite{dh, ag2}, whereas it does not vanish if one considers the exact expression or the higher orders of $B$.

 \section{\label{sec4}Energy extraction from magnetized Kerr spacetime through (magnetic) Penrose process}
 Using the results obtained in Secs. \ref{sec2} and \ref{sec3}, we deduce the efficiency of (magnetic) Penrose process for the magnetized Kerr spacetime in this section.
 For the magnetized Kerr spacetime (Eq. \ref{Kerrm}), one can write
\begin{eqnarray}
 g_{tt}=-\left(\f{\D \S |\L|^2}{A}-\f{A\varpi^2\sin^2\th}{\S |\L|^2}\right)\, , \,\,\,\,\, g_{t \phi}=-\f{\varpi A|\L_0|^2 \sin^2\theta}{\S|\L|^2}
\, , \,\,\,\,\,
  g_{\phi \phi} &=& \f{\varpi A|\L_0|^4 \sin^2\theta}{\S|\L|^2}.
  \label{com}
\end{eqnarray}

Note that $\O_+$ and $\O_-$ achieve the same value at the event horizon ($r=R_h$), i.e., 
\begin{equation}
    \Omega_{h} = \frac{d\phi}{dt}\Bigg|_{r=R_h} = -\frac{g_{t\phi}}{g_{\phi\phi}}\Bigg|_{r=R_h}  =  \frac{a}{2MR_h}\left(\frac{1-a^{2}M^{2}B^{4}}{1+a^{2}M^{2}B^{4}}\right)+\frac{3}{4}\left(\frac{aM^{2}B^{4}}{1+a^{2}M^{2}B^{4}}\right)
\end{equation}
which reduces to $\O_h^{\rm Kerr}=a/(2MR_h)$ \cite{jh} for $B \ra 0$.
Substituting Eq. (\ref{atn}) and Eq. (\ref{com}) in Eq. (\ref{etaf}), one can obtain the exact expression of the efficiency $(\eta_{\rm MPP})$ of MPP at a specific $r$ for the magnetized Kerr spacetime. Setting $q \ra 0$ in $\eta_{\rm MPP}$, one can obtain the exact expression of the efficiency $(\eta_{\rm PP})$ of the PP at a specific $r$ for the magnetized Kerr spacetime. However, in this paper, our aim is to calculate the maximum efficiency which is obtained if the split occurs close to the boundary of the BH (event horizon: $R_h$) and boundary of the SS (:$R_s$). Therefore, we substitute $r \ra R_h$ for a BH and $r \ra R_s$ for a SS in the exact expressions of $\eta_{\rm MPP}$ and $\eta_{\rm PP}$, and obtain the useful expressions and plots in the next section. However, we do not show here the exact expressions of $\eta_{\rm MPP}$ and $\eta_{\rm PP}$, as those are very big in size. \footnote{The exact expressions ($\eta_{\rm MPP}$ and $\eta_{\rm PP}$) of the same can be available upon request.} The expressions of $\eta_{\rm MPP}$ and $\eta_{\rm PP}$ are important only for the numerical calculations, which we use to plot the curves presented in this paper. Thus, we only mention the (approximate) expressions of the efficiency in the next section for the various special cases to understand how the efficiency is affected by the physical parameters in the weak magnetic fields ($B << M^{-1}$). Note that $\eta_{\rm MPP}$ depends on the various physical parameters, such as, mass $(M)$ and Kerr parameter $(a)$ of the collapsed object, the intensity of the magnetic field ($B$) around it and the charge by mass ratio ($\l=|q/m|$) of the outgoing fragment. All of these play an influential role regarding the magnitude and nature of the efficiency. In this paper, we only consider $|q_e/m_e|$ of the electron (i.e., $\l_e=|q_e/m_e|=7.24 \times 10^{21}$) following \cite{dh}, and compare it with the $|q_p/m_p|$ of the proton (i.e., $\l_p=|q_p/m_p|=3.94 \times 10^{18}$) in some interesting cases. Note that almost all the curves in this paper are plotted for $\l_e$, if it is not stated specifically.

\subsection{\label{Smpp}Efficiency of magnetic Penrose process for the magnetized Kerr black hole}

\begin{figure} [!h]
	\begin{minipage} [ch]{0.45\linewidth}
	\hspace*{-1.2cm} \centering \includegraphics [width=1.2\textwidth] {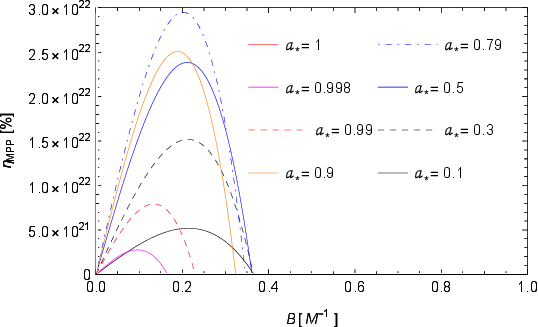} 
	\caption{$\eta_{\rm MPP}$ versus $B$ for the Kerr BHs surrounded by an ultra-strong magnetic field. It shows that the peak ($\eta_{\rm max}$) of $\eta_{\rm MPP}$ starts to decrease for $a_* > 0.79$, and $\eta_{\rm MPP} \leq 20.7\%$ for $a_*=1$, as seen from FIG. \ref{ep}. As the value of $\eta_{\rm MPP}$ is very low for $a_*=1$ compared to the values shown along the Y-axis in the present plot, we draw FIG. \ref{ep} separately. See Sec. \ref{Smpp} for details.}
	\label{ftbh_mpp}
    \end{minipage} \hfill
	\begin{minipage} [ch]{0.45\linewidth}
		\hspace*{-1.2cm} \centering \includegraphics [width=1.1\textwidth] {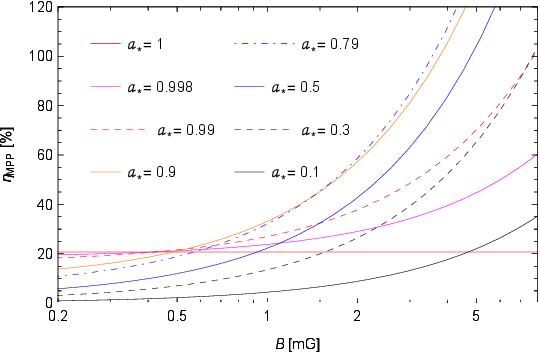} 
		\caption{$\eta_{\rm MPP}$ versus $B$ for the Kerr BH of $M = 10 M_{\odot}$ surrounded by a mG order magnetic field. The solid red line represents $\eta_{\rm MPP} (\approx 20.7\%$) for $a_*=1$. This figure is similar to FIG. 1 of \cite{dh} except the new feature of decreasing $\eta_{\rm MPP}$ for $a_* > 0.79$. See Sec. \ref{Smpp} for details.}
		\label{fabh_mpp}
	\end{minipage} \hfill
\end{figure}

\begin{figure} [!h]
 \begin{minipage} [ch]{0.45\linewidth}
	\hspace*{-1.2cm} \centering \includegraphics [width=1.2\textwidth] {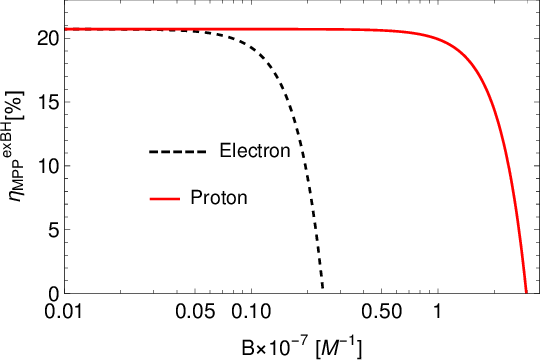} 
	\caption{$\eta_{\rm MPP}^{\rm exBH}$ versus $B$ for an extremal Kerr BH ($a_*=1$) surrounded by a magnetic field of $B \sim 10^{-7}M^{-1}$. It shows that the efficiency for an extremal Kerr BH remains constant (as seen from the solid red line of FIG. \ref{fabh_mpp}), then decreases and vanishes in the comparatively stronger magnetic field shown in FIG. \ref{ftbh_mpp}. See Sec. \ref{Smpp} for details.}
	\label{ep}
\end{minipage} \hfill
 \begin{minipage} [ch]{0.45\linewidth}
	\hspace*{-.6cm}\centering \includegraphics {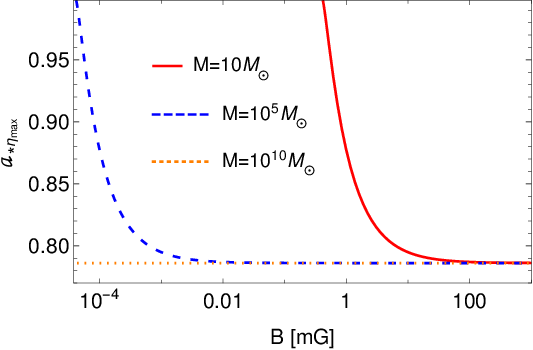}
	\caption {$a_{*\eta_{\rm max}}$ versus $B$ in the weak magnetic field for the different BHs. The orange curve indicates that $\eta_{\rm MPP}(\equiv \eta_{\rm max})$ becomes maximum for $a_*=0.786\approx 0.79$, whereas the upper portion of the curve indicates $\eta_{\rm MPP} < \eta_{\rm max}$. In a similar manner, the red/blue curve stands for $\eta_{\rm max}$ for the respective BHs, whereas the right side of the same curve $\eta_{\rm MPP} < \eta_{\rm max}$. See Sec. \ref{Smpp} for details.}
	\label{fetamax}
	\end{minipage} \hfill
\end{figure}

FIG. \ref{ftbh_mpp} shows the nature of $\eta_{\rm MPP}$ for the various values of $a_*$ and $B$ (in the unit of $M^{-1}$) for \footnote{We do not consider $\l=q/m=1$, as it seems to be unphysical.}  $\l_e = 7.24 \times 10^{21}$.
    As the chosen scale is very large, all the curves seems to be originating from the origin at $(0,0)$. However, in reality, depending on the values of Kerr parameter the curves originate from different values of the Y-axis. This is clear from FIG. \ref{fabh_mpp} which is plotted for the weak magnetic field (mG order). Note that FIG. \ref{fabh_mpp} is exactly similar to FIG. 1 of \cite{dh} except the new feature of decreasing $\eta_{\rm MPP}$ for $a_* > 0.786 \approx 0.79$, which shows that our exact result is consistent with the earlier result (in the weak magnetic field limit: $B << M^{-1}$) obtained in \cite{dh} as the efficiency of MPP.  FIG. \ref{fabh_mpp} shows that $\eta_{\rm MPP}$ can be more than $100\%$ even in the mG order magnetic field. Remarkably, FIG. \ref{ftbh_mpp} shows that the energy extraction could be much higher ($\sim 10^{22}\%$) if the Kerr BH is surrounded by an ultra-strong magnetic field ($B \sim M^{-1}$, see also Eq. \ref{bmax}). The enormous energy which is extracted from the magnetized Kerr BH, may not only come from the rotational energy of the BHs, but it could also come from the energy of the electromagnetic fields. This is because the ultra-strong magnetic field itself could be act as an energy reservoir \cite{leh}, and it can release enormous amount of energy in many cases \cite{leh, cbns}. However, FIG. \ref{ftbh_mpp} reveals that the efficiency increases upto a particular value of $B (\equiv B_p)$ with increasing the magnetic field, attains a peak at $B =B_p$ and then decreases to zero at a particular value of $B$ ($\equiv B_0$). Intriguingly, the efficiency at $B_p$ increases for increasing the value of $a_*$ from $0$ to $a_*=0.786 \approx 0.79$, and, then  
    decreases for  $0.786 < a_* \leq 1$. The efficiency ($\eta_{MPP}^{\rm exBH}$) for the extremal Kerr BH is almost invisible in FIG. \ref{ftbh_mpp}, because it first remains constant at $\eta_{MPP}^{\rm exBH}=20.7\%$ ($<<$ the values of Y-axis in FIG. \ref{ftbh_mpp}), and, then vanishes at $B_0 \sim 2.43 \times 10^{-8}M^{-1}$ (for electron) which is clear from FIG. \ref{ep}. If the outgoing particle is a proton, $B_0 \sim 2.97 \times 10^{-7}M^{-1}$ for the extremal Kerr BH.

Note that the exact expression of the efficiency ($\eta_{\rm MPP}$) of MPP for the extremal magnetized Kerr BH ($a \ra M, r \ra M)$) is deduced as
  \begin{eqnarray}
   \eta^{\rm exBH}_{\rm MPP} = \f{1}{4}\left[\f{(8+8B^2M^2+8B^4M^4+B^8M^8)^{1/2}}{1+B^2M^2} -2 \right] - \f{q B^3M^3}{2m}.\f{(4+7B^2M^2)}{(1+B^2M^2)}.
   \label{exbhmpp}
  \end{eqnarray}
  For $BM << 1$, if one considers upto the linear order in $B$ of Eq. (\ref{exbhmpp}), the efficiency becomes exactly the same ($20.7\%$) to the unmagnetized extremal Kerr BH case, as predicted in \cite{dh}, and also seen from Eq. (\ref{mppbh}).
   For $BM=1$, the efficiency (Eq. \ref{exbhmpp}) would be 
\begin{eqnarray}
 \eta^{\rm exBH}_{\rm MPP}|_{BM \ra 1}=\f{1}{8}-\f{11q}{4m}.
 \label{bm1}
\end{eqnarray}
Eq. (\ref{bm1}) reveals that the efficiency becomes negative for $q/m > 1/22$ which is much less than $\l_e$ and $\l_p$. Thus, it is almost impossible to extract the BH energy through MPP in this special case. 

The similar pattern in the efficiency curves of FIG. \ref{ftbh_mpp} in the mG order magnetic field can also be noticed in FIG. \ref{fabh_mpp}, i.e., $\eta_{\rm MPP}$ increases with the increasing the value of $B$ and $0 < a_* < 0.786$, but decreases for $0.786 < a_* \leq 1$. The efficiency becomes exactly same to the efficiency of the regular extremal Kerr BH, i.e., $20.7\%$ which is represented by the solid red line in FIG. \ref{fabh_mpp} even in the presence of magnetic fields. This
is the well-known gravitational analogue of Meissner effect in
superconductor, where a conductor turns to a superconductor and all fields are expelled out \cite{dh}. Therefore, the effect of magnetic fields vanishes (due to $R_h \ra M$) from the dominating second term (linear order in $B$) of the following efficiency expression, 
 \begin{eqnarray}\nonumber
  \eta_{\rm MPP}^{\rm BH} = \f{1}{2}\left(\sqrt{\f{2M}{R_h}}-1 \right)+\f{q}{m}aB \left(1-\f{M}{R_h} \right)- \f{a^2B^2}{2\sqrt{2}}\left(\f{M}{R_h}\right)^{3/2}+\f{q}{m}\f{aMB^3}{R_h^2} \left[4Ma^2+R_h(a^2-7M^2)\right] +\mathcal{O}(B^4).
  \\
  \label{mppbh}
 \end{eqnarray}
 Eq. (\ref{mppbh}) is similar to Eq. (9) of \cite{dh}, if one considers upto the terms linear order in $B$. Note that we obtain Eq. (\ref{mppbh}) from our exact expression of $\eta_{\rm MPP}$ which reduces to Eq. (\ref{mppbh}) for $BM << 1$ and $r \ra R_h$. The term $q/m$ which is responsible for the MPP, appears for the odd orders in $B$. This is clear from the second and fourth terms of Eq. (\ref{mppbh}). Interestingly, the gravitational analogue of the Meissner effect can diminish the efficiency to zero for the further increment of the magnetic field for $a_*=1$, which is clear from FIG. \ref{ep} as well as the third term of Eq. (\ref{mppbh}). This does not reflect in Eq. (9) of \cite{dh}, as they considered the terms linear in $B$. Hence, we obtain the additional contribution shown in FIG. \ref{ep} in terms of the gravitational Meissner effect. One should note here that the BH slowly starts to expel the magnetic field around $a_* \sim 0.786$ and it is fully expelled for $a_*=1$. That is why, the efficiency of MPP starts to decrease from $a_*=0.786$ and is continued until $a_*=1$ which is clear from FIGs. \ref{ftbh_mpp}--\ref{ep}. 
 
A close observation of FIG. \ref{fetamax} reveals that $\eta_{\max}$ for $a_{*\eta_{\max}} \approx 0.786$, and the decreasing trend of $\eta_{\rm MPP}$ for the range $0.786 < a_* \leq 1$ remains unchanged for the BH mass $M \gtrsim 10M_{\odot}$ and its surrounding magnetic fields $B > 10$ mG including the ultra-strong magnetic fields (see FIGs. \ref{ftbh_mpp} and FIG. \ref{ep}). For $B < 10$ mG, the value of $a_{*\eta_{\max}}$ depends on the mass of the BH as seen from the solid red curve of FIG. \ref{fetamax}. Similarly, it also holds for the BHs of $M > 10^5M_{\odot}$ surrounded by a $B > 1\mu$G, represented by the dashed blue curve in FIG. \ref{fetamax}. The dotted orange curve of the same (FIG. \ref{fetamax}) shows that $a_{*\eta_{\max}} \approx 0.786$ and the range remains almost unchanged for a supermassive BH of $M=10^{10}M_{\odot}$ even in the extremely weak magnetic field. The orange curve indicates that $\eta_{\rm MPP}(\equiv \eta_{\rm max})$ becomes maximum for $a_*=0.786$, whereas the upper portion of the curve indicates $\eta_{\rm MPP} < \eta_{\rm max}$. Similarly, the range ($0.786 < a_* \leq 1$) remains unchanged for a BH of $M=10^5M_{\odot}$ with $B > 0.01$ mG, but the range decreases (e.g., $0.87 < a_* \leq 1$ valid for $B > 10^{-4}$ mG) if the magnetic field decreases further. This can be seen from the dashed blue curve of FIG. \ref{fetamax}. A particular point in a specific curve of FIG. \ref{fetamax} indicates that $\eta_{\rm MPP}$ becomes maximum for that specific value of  $a_* \equiv a_{*\eta_{\rm max}}$ and $B$. For $a_* > a_{*\eta_{\rm max}}$ for that particular value of magnetic field, $\eta_{\rm MPP} < \eta_{\rm max}$. The value of $a_{*\eta_{\rm max}}$ in the weak magnetic field can be obtained by setting $\p \eta_{\rm MPP}^{\rm BH}/\p a=0$ for the terms linear in $B$ of Eq. (\ref{mppbh}), and solve for $a$. Note that if the outgoing particle is considered as a  proton, the value $a_{*\eta_{\max}} \approx 0.786$ remains same, but the curves of FIG. \ref{fetamax} shifts toward right (i.e., away from the Y-axis). For example, the red curve of FIG. \ref{fetamax} starts to move upward at a value $B \sim 10^4$ mG instead of $B \sim 10$ mG.

\begin{figure} [!h]
\centering \includegraphics [width=.6\textwidth] {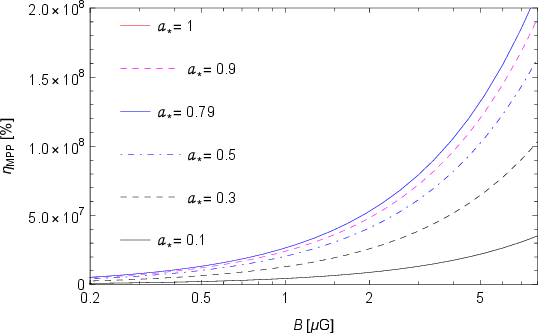} 
	\caption{$\eta_{\rm MPP}$ versus $B$ for a Kerr BH of $M = 10^{10} M_{\odot}$ surrounded by a $\mu$G order magnetic field. $\eta_{\rm MPP}$ for $a_*=1$ is so small $(\approx 20.7\%$) compared to the values mentioned along the Y-axis, that it cannot be seen in this plot (see FIG. \ref{ep}). The feature of all curves is qualitatively similar to FIG. \ref{fabh_mpp} including $a_{*\eta_{\rm max}} \approx 0.79$ (see FIG. \ref{fetamax}). See Sec. \ref{Smpp} for details.}
	\label{fabh_mpp_mu}
\end{figure}

 Note that all of the above-mentioned conclusions remain unchanged if we calculate the efficiency for a supermassive BH (SMBH) of mass $M=10^{10}M_{\odot}$ which surrounded by a $\mu$G order magnetic field. For example, FIG. \ref{fabh_mpp_mu} shows that the efficiency of MPP is much greater than $100\%$, and an enormous amount of energy can be extracted from that SMBH. This could explain why the AGNs are so luminous during their active phase.

\subsection{\label{Spp}Efficiency of Penrose process for the magnetized Kerr black hole}
\begin{figure} [!h]
		\centering \includegraphics {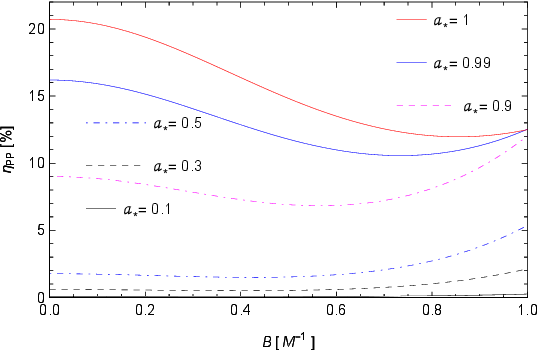}
        \caption {$\eta_{\rm PP}$ versus $B$ for the Kerr BHs surrounded by an ultra-strong magnetic field. Very close to $B \ra M^{-1}$, the value of $\eta_{\rm PP}$ of the following range: $0.96 < a_* \lesssim 0.99$ increases, which is slightly greater than $\eta^{\rm exBH}_{\rm PP}=12.5\%$. For $0.99 < a_* \leq 1$, the efficiency decreases. See Sec. \ref{Spp} for details.}
        \label{ftbh_pp}
\end{figure}

As we mentioned in Sec. \ref{sec4}, one can easily obtain the exact expression of $\eta_{\rm PP}$ of the ordinary PP for the magnetized Kerr BH, if $q \ra 0$ and $r \ra R_h$ are substituted in the exact expression of MPP. For the weak magnetic field, the even order terms (e.g., first and third terms) of $B$ of Eq. (\ref{mppbh}) can be considered for the approximate expression of efficiency for the ordinary PP, i.e.,
  \begin{eqnarray}
  \eta_{\rm PP}^{\rm BH} = \f{1}{2}\left(\sqrt{\f{2M}{R_h}}-1 \right)- \f{a^2B^2}{2\sqrt{2}}\left(\f{M}{R_h}\right)^{3/2}+\mathcal{O}(B^4).
  \label{ppbh}
 \end{eqnarray}
 Although the two terms of Eq. (\ref{ppbh}) are purely geometric, the second term is responsible for decreasing the efficiency of PP with increasing the magnetic field satisfied by $BM << 1$. This could be clear from the starting part of the curves of FIG. \ref{ftbh_pp}. 
However, the curves of this figure are plotted with the exact expression of $\eta_{\rm PP}$. It indicates that $\eta_{\rm PP}^{\rm BH}$ decreases with increasing the magnetic field, but it can increase further in the ultra-strong magnetic field depending on the value of Kerr parameter which is clear from FIG. \ref{ftbh_pp}.  

In case of the ordinary Penrose process for the magnetized extremal Kerr BH, we obtain the exact expression of efficiency as
\begin{eqnarray}
   \eta^{\rm exBH}_{\rm PP} = \f{1}{4}\left[\f{(8+8B^2M^2+8B^4M^4+B^8M^8)^{1/2}}{1+B^2M^2} -2 \right]
   \label{exbhpp}
  \end{eqnarray}
  by substituting $q \ra 0$ in Eq. (\ref{exbhmpp}).
The above equation shows that the efficiency of the extremal Kerr BH decreases from $20.7\%$ to $1/8 \sim 12.5\%$ for the ordinary Penrose process in the presence of strong magnetic field.  The minimum efficiency ($\sim 12\%$) in this case is obtained for $B=(\sqrt{3}-1)^{1/2}M^{-1} \approx 0.8556M^{-1}$. Intriguingly, very close to $B \ra M^{-1}$, the value of $\eta_{\rm PP}$ of the following range: $0.96 < a_* \lesssim 0.99$ increases, which is slightly greater than even $\eta^{\rm exBH}_{\rm PP}=12.5\%$. However, the efficiency decreases for $0.99 < a_* \leq 1$. This could also be an indication of the trailing part of gravitational Meissner effect which is similar to the trailing part of the curves shown in FIG. \ref{ep}. This feature is not very clear from FIG. \ref{ftbh_pp}. Although he red and blue solid curves of FIG. \ref{ftbh_pp} are seem to be overlapped, in relaity, the blue curve crosses the red curve close to $B \ra M^{-1}$. If one extends the X-axis further, the feature would be visible.

 \subsection{\label{Sm}Efficiency of magnetic Penrose process for the magnetized Kerr superspinar}
 
  In case of the magnetized Kerr SS, a huge energy extraction could be possible from the vicinity of the boundary of the superspinar even in the absence of the magnetic field. Note that the energy extraction from $r \ra 0$ (very close to the singularity) is not feasible. The quantum gravity effects should be important in the region where the spacetime curvature approaches the Planck scale, i.e., very close to the ring singularity \cite{bambif}. It is also widely accepted that quantum gravity will resolve the singularity resulting in an overspinning object with a boundary at a positive value of $r$, which is referred to as a `superspinar' \cite{gh, ckp}. Therefore, we restrict our probe for $ r \geq 0$ \cite{ckp}, and the limit of the point of split is considered at the boundary ($R_s$) of the SS: $R_s=10^{-3}M$ following \cite{bambi}, as already discussed in Sec. \ref{sec2}. This radius ($R_s$) is still conservative {\footnote{The value of $R_s$ could be upto the Planck length, i.e., $R_s > 1.6 \times 10^{-35}$ meter.}}  in this aspect that the curvature of the spacetime as an astrophysical object is tiny (in the Planck unit) at the distance $10^{-3}M$ from $r=0$ \cite{bambi}. Thus, $r$ could be replaced by $R_s$ in the exact expression of $\eta_{\rm MPP}$ (as mentioned in Sec. \ref{sec4}) to calculate the efficiency of MPP in case of the Kerr SS.

\begin{figure} [!h]
	\begin{minipage} [ch]{0.45\linewidth}
		\hspace*{-1.2cm} \centering \includegraphics [width=1.2\textwidth] {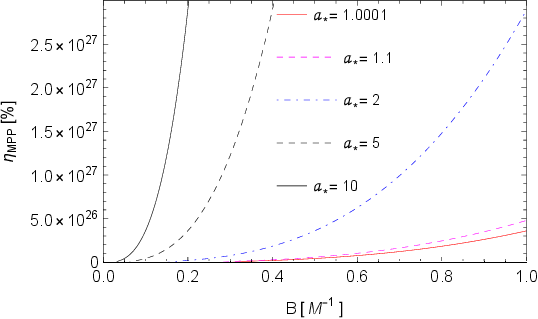} 
		\caption{$\eta_{\rm MPP}$ versus $B$ for the Kerr SSs surrounded by an ultra-strong magnetic field. Although $\eta_{\rm MPP}$ increases with both $B$ and $a_*$ in this figure, very close to the Y-axis ($B << M^{-1}$) it behaves in a opposite way, which is depicted separately in FIG. \ref{fass_mpp}. See Sec. \ref{Sm} for details.}
		\label{ftss_mpp}
	\end{minipage} \hfill
	\begin{minipage} [ch]{0.45\linewidth}
		\hspace*{-0.8cm}\centering \includegraphics {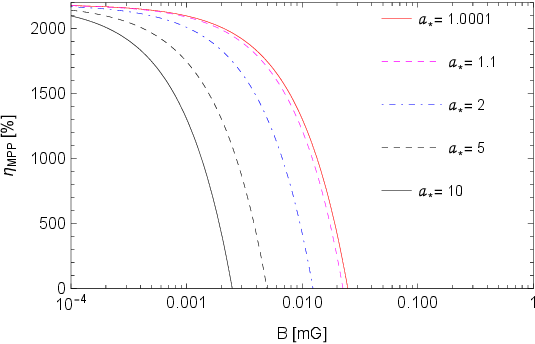}
		\caption {$\eta_{\rm MPP}$ versus $B$ for a Kerr SS of $M = 10 M_{\odot} $ surrounded by a mG order magnetic field. It shows that $\eta_{\rm MPP}$ decreases with both $B$ and $a_*$, which is opposite (see FIG. \ref{ftss_mpp}) to the behavior in the ultra-strong magnetic field. See Sec. \ref{Sm} for details.}
		\label{fass_mpp}
	\end{minipage}\hfill
\end{figure}

For the weak magnetic fields ($BM << 1$), the exact expression of $\eta_{\rm MPP}$ reduces to
 \begin{eqnarray}\nonumber
  \eta_{\rm MPP}^{\rm SS} = \f{1}{2}\left(\sqrt{\f{2M}{R_s}}-1 \right)+\f{q}{m}aB \left(1-\f{M}{R_s} \right)- \f{B^2}{8\sqrt{2MR_s^3}}
 \left[R_s^3 (R_s-2 M) + a^2 (4 M^2 + R_s^2)\right]
 \\
  +\f{q}{m}\f{aMB^3}{2R_s^2} \left[a^2 (M+R_s)-R_s^2(R_s+5M)\right] +\mathcal{O}(B^4)
  \label{mppss}
 \end{eqnarray}
 for $r \ra R_s$. The efficiency profile plotted using the exact expression of $\eta_{\rm MPP}$ is given in FIG. \ref{ftss_mpp} and FIG. \ref{fass_mpp} for the ultra-strong and mG order magnetic fields, respectively. In the case of a SS surrounded by an ultra-strong magnetic field, FIG. \ref{ftss_mpp} shows a rather deceiving nature of a monotonic increase in efficiency with increasing the value of $a_*$. On initial inspection, besides extremely high $\eta_{\rm MPP}$ value nothing seems that different, but by further zooming into the origin we observe a rather interesting feature of the plot (see FIG. \ref{fass_mpp}). It shows that $\eta_{\rm MPP}$ suffers a drastic drop for a very small increment in the magnetic field, e.g., although $\eta_{\rm MPP} \approx 2186\%$ (see Eq. \ref{mppss}) for $B \ra 0$, it vanishes at $B \sim 0.005$ mG for $a_* = 5$. 
FIG. \ref{fass_mpp} reveals that 
 the initial rate (in the mG order magnetic field) of dropping in efficiency with higher value of Kerr parameter is higher, whereas the rising rate of efficiency with higher value of Kerr parameter is also higher in the ultra-strong magnetic fields (see FIG. \ref{ftss_mpp}).
 As seen from FIG. \ref{ftss_mpp}, the rise in $\eta_{\rm MPP}$ is observed in the ultra-strong magnetic field at which a very large efficiency could be obtained. The MPP efficiency of the magnetized Kerr SS is larger (FIG. \ref{ftss_mpp}) than the efficiency attained by a magnetized Kerr BH (FIG. \ref{ftbh_mpp}). Moreover, the feature of the curves of FIG. \ref{ftss_mpp} is completely different form the feature of the curves of FIG. \ref{ftbh_mpp}.

\subsection{\label{Sp}Efficiency of Penrose process for the magnetized Kerr superspinar}

\begin{figure} [!h]	
\centering \includegraphics[width=.6\textwidth]{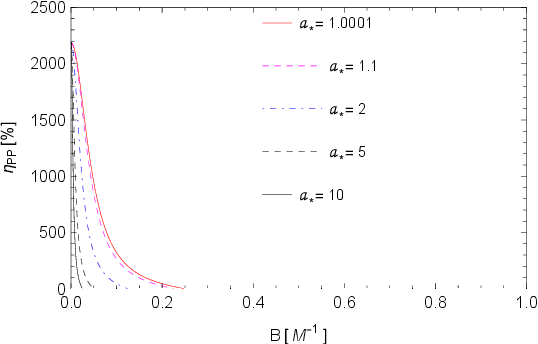}
		\caption {$\eta_{\rm PP}$ versus $B$ for the Kerr SSs surrounded by an ultra-strong magnetic field.  It shows that $\eta_{\rm PP}$ decreases with both $B$ and $a_*$, the feature of the curves are opposite to the curves of FIG. \ref{ftss_mpp}. See Sec. \ref{Sp} for details.}
		\label{ftss_pp}
\end{figure}

The efficiency profile of the ordinary PP for the magnetized Kerr SS is plotted using the exact expression of $\eta_{\rm PP}$, and is given in FIG. \ref{ftss_pp}.

In the weak magnetic field ($BM << 1$), one can obtain the efficiency of PP as
 \begin{eqnarray}
  \eta_{\rm PP}^{\rm SS} = \f{1}{2}\left(\sqrt{\f{2M}{R_s}}-1 \right)- \f{B^2}{8\sqrt{2MR_s^3}}
 \left[R_s^3 (R_s-2 M) + a^2 (4 M^2 + R_s^2)\right]+\mathcal{O}(B^4).
  \label{ppss}
 \end{eqnarray}
 by substituting $q \ra 0$ in Eq. (\ref{mppss}), 

 In case of the ordinary (unmagnetized) Kerr superspinar (i.e., $B \ra 0$), the exact expression of $\eta_{\rm MPP}$ reduces to
  \begin{eqnarray}
   \eta_{\rm PP}^{\rm SS}|_{B\ra 0} = \f{1}{2}\left(\sqrt{\f{2M}{R_s}}-1 \right)
  \end{eqnarray}
  that does not depend on the value of mass and spin parameter of SS. It only depends on the value of $R_s/M$. For example, $\eta_{\rm PP}^{\rm SS}|_{B\ra 0} \approx 2186\%$ for $R_s=10^{-3}M$. This shows that a huge energy extraction could be possible from an ordinary Kerr SS through the ordinary PP, even in the absence of magnetic field.

   This suggest that the SSs are also likely to be very luminous
during their active phase \cite{gh} like BHs. This could explain why the AGNs are so luminous \cite{gh}. In case of the SSs, the ergoregion fills a torus \cite{ckj} with an opening angle \cite{ckj, gh} along the axis of rotation for $a_* > 1$. This phenomenon could facilitate the formation of relativistic jets. The particles falling in from the innermost stable circular orbit (ISCO) are trapped by the gravitational well. This produces so much high pressure in the central region that the natural escape route for the particles along the rotation axis as they cannot overcome the frame dragging of the ergoregion \cite{gh}. As the accretion disks are full of charged particles, a magnetized accretion disk with a weak magnetic field (e.g., $1~\mu$G) could be able to explain the relativistic jets from the view of MPP and PP in the case of Kerr SS.
\\

\begin{figure}[h!]
 \begin{center}
\subfigure[BH: $a_*=0.3$. Red curve indicates $\eta_{\rm MPP} \sim 0.6\%$ with $\l_p$.]{\includegraphics[width=3.1in,angle=0]
{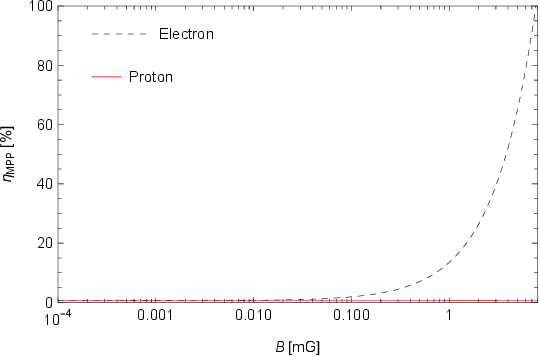}}
\hspace{0.05\textwidth}
\subfigure[SS: $a_*=2$]{\includegraphics[width=3.1in,angle=0]{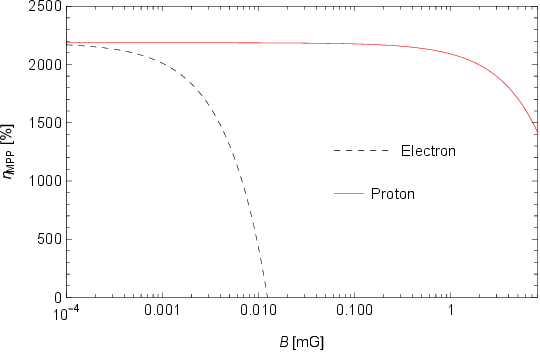}}
\caption{\label{fcep}$\eta_{\rm MPP}$ versus $B$ with different $\l$ is of the order of a particle like electron and a proton in the vicinity of two different Kerr collapsed objects (BH and SS) with $M = 10 M_{\odot}$. See Sec. \ref{Sp} for details.}
\end{center}
\end{figure}

Now lets put forth some efficiency ($\eta_{\rm MPP}$) comparisons for the particles where magnitude of $q/m$ resembles that of particles such as electrons and protons. Panel (a) of FIG. \ref{fcep} shows that, the particles with higher $q/m$ are much more efficient as compared to the particles with lower $q/m$ ($\eta_{\rm MPP} \approx 0.6\%$), in the case of magnetized Kerr BH. In the case of magnetized Kerr SS, the particles with higher $q/m$ suffers a sudden efficiency drop with increase in magnetic field and possesses comparatively lower efficiency than the particles with lower $q/m$ value. In addition to having higher efficiency, they can hold onto that efficiency for longer ranges of magnetic field. So a SS can accelerate a higher mass particle much more efficiently than a comparatively lower mass one and a BH can accelerate a lower mass particle much more efficiently than a heavier one, for the given range of magnetic fields.

\section{\label{sec6}Summary and discussion} 
Several important and interesting results have been presented in this paper. Below we summarize it as follows:
\begin{enumerate}
 \item Considering the exact solution of the magnetized Kerr spacetime, we have derived the exact expression of MPP and PP which are valid for the magnetized Kerr BH as well as magnetized Kerr SS.
 
 \item Our result is applicable from the weak magnetic field to the ultra-strong magnetic field (i.e. $0$ to any arbitrary value of $B$) which can even distort the original Kerr geometry. Unlike the earlier works, we have not assumed the weak-field approximation in $B$, and still our results are consistent with the previous studies carried out with the similar motivation. 
 
 \item We have shown that although the efficiency of MPP increases (much more than $100\%$) upto a certain value of ultra-strong magnetic field ($B_p$), it decreases to zero after crossing that particular value ($B_p$), in case of the magnetized Kerr BHs. This indicates that a ultra-strong magnetic
field could resist the extraction of energy from a Kerr BH.
 On the other hand, $\eta_{\rm MPP}$ shows the monotonically increasing behavior in case of the magnetized Kerr SSs in the ultra-strong magnetic fields. 
 
 \item One intriguing feature that emerges is, $\eta_{\rm MPP}$ acquires the maximum value ($\sim 10^{22}\%$) for $a_* \approx 0.786$ (unlike $a_*=1$ which occurs for the ordinary PP in the unmagnetized Kerr BH case), and $\eta_{\rm MPP}$ decreases for the range $0.786 < a_* \leq 1$, if the Kerr BH surrounded by an ultra-strong magnetic field. This is also true for the weak magnetic fields (see FIG. \ref{fabh_mpp}), but there is some limitations as discussed in Sec. \ref{Smpp}. This indicates that the BH starts to expel the magnetic field around $a_* \sim 0.786$ and it is fully expelled for $a_*=1$ due to the gravitational Meissner effect. 
 
 \item MPP efficiency for the extremal Kerr BH was earlier predicted to be remain constant ($20.7\%$) with $B$, as it was obtained only upto the linear order in magnetic field $B$. From our exact calculation, it is shown that the MPP efficiency remains constant upto a specific value of $B$ (see FIG. \ref{ep}). After that, the MPP efficiency decreases to zero. This is an additional effect on top of the gravitational Meissner effect.
 
 \item For the first time, we provide an equivalent mechanism of MPP/PP by which the energy from a Kerr SS (if it exists in nature) could be extracted. To the best of our knowledge, the magnetic Penrose process for a Kerr SS has not been formulated so far.

\item We show that a particle with higher $q/m$ (charge to mass ratio) is much more efficient as compared to the particle with lower $q/m$, in case of the MPP for a magnetized Kerr BH. In case of the MPP for a magnetized Kerr SS, the particles with higher $q/m$ suffer a sudden efficiency drop with increase in magnetic field and possesses comparatively lower efficiency than the particles with lower $q/m$ value. 

\item As a special case of MPP, we have also derived the exact expression of the ordinary PP for the magnetized Kerr spacetime. We have shown that the efficiency of PP for the magnetized Kerr BHs decreases with increasing the magnetic field, but it can slightly increase close to $B \sim M^{-1}$ depending on the value of Kerr parameter. In case of the Kerr SS (with $R_s \sim 10^{-3}M$), the efficiency of PP decreases from $2186\%$ to $0$ for increasing the value of magnetic field from $0$ to some ultra-strong magnetic field. 

\item Based on our results we also conclude that for a given mass of a collapsed object, a Kerr SS provides higher $\eta_{\rm MPP}$ comparing to a Kerr BH in the ultra-strong magnetic field. In a similar manner, a Kerr SS provides higher $\eta_{\rm PP}$ comparing to a Kerr BH, i.e., $\eta_{\rm PP}^{\rm SS} > \eta_{\rm PP}^{\rm BH}$ for a mG order magnetic field. Note that the Kerr SS has not been detected yet.

\item In the ultra-strong magnetic field, it is almost impossible to extract the energy from a Kerr BH through MPP (for $B > 0.36M^{-1}$) and from a Kerr SS through PP (for $B > 0.25M^{-1}$). Thus, an ultra-strong magnetic field could stop the energy extraction from the Kerr collapsed objects. In these cases, the magnetic fields could act as a shield \cite{shld}, and does not produce outflows from the active galactic nuclei (AGN) in the form of astrophysical jets. In other words, it may prevent any charged and neutral particles \cite{gp} escaping to infinity, and, thereby, refrain to produce any outflows from a collapsed object.

\item If a $B \sim 10$ mG can help to reach $\eta_{\rm MPP} > 100\%$ (as shown in \cite{dh}), it is not very surprising to reach $\eta_{\rm MPP} \sim 10^{22}\%$ for a $B \sim M^{-1}$ (see Eq. \ref{bmax}).
This enormous energy which is extracted from the magnetized Kerr spacetimes, does not only come from the rotational energy of the collapsed objects, but it could also come from the surrounded electromagnetic fields. For instance, the second term of Eq. (\ref{mppbh}) consists of $qB/m$ ($\equiv \o_L$ which is related to the Larmor precession \cite{fls}) multiplied by $a$ (spin parameter of the Kerr BH). The coupling between a very large value of $\o_L$ in the ultra-strong magnetic field and the spin of the BH could play an important role for producing such an enormous energy for MPP. It could also be useful to note here that a strong magnetic field could be act as an energy reservoir \cite{leh}, and it can release enormous amount of energy in many cases \cite{leh, Tursunov20,cbns}. 
\end{enumerate}

Although the upper limit of the magnetic fields in the X-ray corona of the BH Cygnus X-1 rises to $10^7$ G \cite{santo}, emission models of Sagittarius A* and M87* provide $B \sim 30-100$ G \cite{eatough} and $B \sim 1-30$ G \cite{eht7} respectively, the ultra-strong magnetic fields ($\sim$ Eq. \ref{bmax}) around a BH has not been detected yet. It is still unclear whether such a ultra-strong magnetic fields ($B > 0.36M^{-1}$) exist in the Universe. If the ultra-strong magnetic field even exists around a collapsed object, it may not be detectable due to the magnetic shielding \cite{shld}.  However, our result could be verified if such a ultra-strong magnetic field is discovered around a collapsed object in future.

The ordinary PP is not enough for its astrophysical viability for Kerr BH (even in the presence of magnetic field), as its efficiency is very low ($\eta_{\rm PP} < 20.7\%$). On the other hand, based on the efficiency produced for a low value of magnetic fields, it is clear that MPP is a astrophysically viable mechanism for producing high energy particles in the weak (e.g., mG, $\mu$G order) magnetic fields. Recent astronomical observations suggests that a collapsed object is generally surrounded by the radiative matters, ionized particles, hot gases and all of these together around the same object constitutes the accretion disc, and this could create a magnetic habitat around it. Blandford and Znajek proposed a mechanism \cite{bz} to extract the rotational energy of a BH through the electromagnetic interaction. The BZ mechanism requires $B \gtrsim 10^5$G for it to be even operative. For example, the maximal BZ power was calculated \cite{dh} as $1.7 \times 10 ^{46}$ erg/s for $M = 10^9 M_{\odot}$ and $B = 10^4$ G. On the other hand, a mG order magnetic field can generate enormous energy from a Kerr collapsed object through MPP. It indicates that the MPP is much more efficient than the BZ mechanism \cite{dh}. Therefore, in recent years serious efforts have been made to directly explain the origin of jets \cite{tch,nar} and ultrahigh-energy cosmic rays \cite{Tursunov20} using MPP. 

The production mechanisms of ultrahigh-energy (e.g. EeV order) cosmic rays which we receive on Earth from the outer space, remains unclear. Almost all the currently available mechanisms are based on the electromagnetic interaction between the accelerated charged particles. Intriguingly, the highest energy of a detected cosmic ray is $3\times 10^{20}$ eV \cite{nasa}. Our result shows that $\eta_{\rm MPP}$ could reach maximum upto $\sim 3 \times 10^{22}\%$ (note that the number is independent of the mass of the BH; see FIG. \ref{ftbh_mpp}) in case of a magnetized Kerr BH and much more for a magnetized Kerr SS. For example, FIG. \ref{ftbh_mpp} shows that a charged particle of $1$ eV could escape to infinity with the energy $10^{20}$ eV from a magnetized Kerr BH of $a_* \sim 0.79$, which is surrounded by $B \sim 0.2M^{-1}$. Thus, the existence of any ultra-strong magnetized BHs in reality \footnote{It is not realistic to expect such a high magnetic field ($B \sim 0.2M^{-1}$) around a SMBH as of now, because $B \sim 0.2M^{-1}$ is equivalent to $10^{10}$ G for a SMBH of $M \sim 10^9M_{\odot}$, and such a strong magnetic field around a SMBH has not been detected yet.} could have explained the production mechanism of EeV range energy cosmic rays.

A ultra-strong magnetic field in the order of $10^{20}$ G may be originated (see \cite{gru, grasso} and references therein) in the early Universe, and the primordial black holes (PBHs) could be immersed in that magnetic fields. Our exact result of MPP in the presence of ultra-strong magnetic fields could be applicable to those strongly magnetized PBHs. Our result could also be important to those BHs which are formed by collapse induced  \footnote{These BHs are generally known as the transmuted BHs \cite{dasg}. If a BH is formed by collapse induced by an endoparasitic BH through capturing of dark
matter in a magnetar of $M \sim 2M_{\odot}$, one may expect a comparable strong magnetic field ($B \sim 0.2M^{-1}$ equivalent to $10^{18}$ G in this case) around that BH, or the BH formed by merging of several such strongly magnetized progenitors. Note that such a strongly-magnetized BH and/or a near-solar mass BH has not been detected yet (but, see \cite{dasg, gw1, gw2}).} by an endoparasitic BH through capturing of dark
matter into a neutron star \cite{gold, dasg} (or, white dwarf \cite{cbns}, magnetar etc.) with magnetic field, and by merging of a BH with one (or multiple) magnetized neutron star(s) or magnetar(s) \cite{lyu83, east, shld}. In such cases, the magnetic field could be prevented from sliding off the newly-formed BH as shown in \cite{lyu, lyu83} (but see \cite{brp}), and one can calculate $\eta_{\rm MPP}$ of such magnetized BHs by using our exact expression of MPP.

As there is a huge discrepancy in the MPP efficiency of Kerr BH and SS for the similar value of magnetic fields, one could distinguish between them by studying the power output from a collapsed object, if the Kerr SS, in fact, exists in nature. The SS and BH could also be distinguished theoretically based on another parameter and that is the energy carried by the lighter particle (e.g., electron) as compared to that of heavier particle (e.g., proton). In fact, if and when a split occurs in ergoregion producing oppositely charged particles, the electrons are generally accelerated outward \cite{Rueda2023} and come out from the polar region \cite{Rueda2023, Tursunov20}. Thus, we have discussed our result focusing on the efficiency of MPP using electron.

Although the original magnetized Kerr metric dealt with in this paper is not asymptotically flat and the magnetic field is considered to be asymptotically uniform following \cite{dh}, it is an exact electrovac solution of Einstein-Maxwell equation. This is also a standard practice to consider a uniform magnetic field around a collapsed object (BH or SS) to explain the several astrophysical phenomena (e.g. see \cite{shi, za, za2, ssb, Tursunov20, nia, dh, c22, ccb} and so on), as it is easier to handle. The lacking of the proper measurements of the exact shapes of the magnetic field configurations \cite{mei, pun} around a collapsed object is also the another reason for assuming the uniform magnetic fields in this paper. Finally, the MPP formalism could be improved by considering the varying magnetic field, which could be more realistic.
\\

{\bf Acknowledgements :} We thank the referees for constructive comments that helped to improve the manuscript from the observational point of view. One of us (CC) thanks M. Astorino for useful discussions regarding the magnetized Kerr spacetime. CC also thanks S. V. Dhurandhar for an useful discussion on the magnetic Penrose process. CC also thanks the Inter-University Centre for Astronomy and Astrophysics (IUCAA), Pune, India, as the part of this work was done during the academic visit to IUCAA under the visiting associateship programme. CC dedicates this paper to his mother, Shree Snigdha Chakraborty who was his constant inspiration in all fields of life, and unfortunately passed away on September 11, 2023.

\end{document}